\documentstyle[12pt,epsf]{article}
\begin {document}

\begin{center}
             {\bf Perturbative Expansion in the Galilean Invariant}\\

               {\bf Spin One-Half Chern-Simons Field Theory}\\

                                C. R. Hagen\\

                    Department of Physics and Astronomy\\
                          University of Rochester\\
                           Rochester, N.Y. 14627\\

\end{center}
\begin{abstract}                                  

     A Galilean Chern-Simons field theory is formulated for the case of two
interacting spin-1/2 fields of distinct masses $M$ and $M^\prime$.
A method for the construction of states containing $N$ particles of mass 
$M$ and $N^\prime$ particles of mass $M^\prime$
is given which is subsequently used to display equivalence to the
spin-1/2 Aharonov-Bohm effect in the $N = N^\prime =1$ sector of the model. 
The latter is then studied in perturbation theory to determine whether there are
divergences in the fourth order (one loop) diagram.  It is found that the
contribution of that order is finite (and vanishing) for the case of parallel
spin projections while the antiparallel case displays divergences which are
known to characterize the spin zero case in field theory as well as in quantum
mechanics.   

\end{abstract}

\bigskip

\noindent I. Introduction

     The Aharonov-Bohm (AB) effect [1] has been studied extensively in recent
years both in the context of quantum mechanics as well as in quantum field
theory.  As an application of wave mechanics it is customarily idealized to the
discussion of the scattering of charged particles from a magnetized filament of
arbitrarily small radius.  Since the exterior of such a filament is a field free
region, there can be no classical force on the particles.  The fact that a
nontrivial scattering cross section is found thus provides a forceful
demonstration of the significance of the vector potential in the quantum
mechanical description of scattering.    

     Although not used in the original AB work the partial wave description of
this phenomenon is of considerable interest.  For the partial wave $f_m(r)$
the relevant equation is

$$\left [{1\over r}{\partial\over\partial r}r{\partial\over\partial r} +k^2 -
{(m+\alpha)^2\over r^2}\right]f_m(r)=0$$

\noindent where $\alpha$ is the flux parameter and $k^2=2ME$ with $M$ the
particle mass and E its nonrelativistic energy.  Standard techniques allow one 
to obtain that the phase shifts $\delta_m$ are given by 

$$\delta_m=-{\pi \over 2}\mid m+\alpha \mid +{\pi \over 2} \mid m \mid$$
     
\noindent so that for small $\alpha$ one finds the nonanalytic form

     $$\delta_0 = -\mid \alpha \mid {\pi \over 2}.$$

\noindent Since in the $m=0$
 partial wave the potential is proportional to $\alpha^2$, this
suggests that a perturbative 
approach may encounter considerable difficulty [2]. 
Aharonov {\em et al}. demonstrated the existence of singularities in such an
expansion using as a model an impenetrable solenoid of finite radius R[3]. 
Although the limit $R\rightarrow 0$ was found to yield the usual AB scattering
amplitude, to any finite order in $\alpha$ the solution consisted of a
complicated expansion in powers of $\alpha \ell n(kR/2)$.  

     Similar studies have been carried out in the context of field theory.  One
prerequisite to such a study was the construction of the so-called pure
Chern-Simons (or photonless) gauge theory as carried out by this writer[4].  It
was subsequently shown [5] that the Galilean limit of such a theory allows one
to formulate what is the only Galilean invariant gauge theory known at this
time.  It is in fact the field theory of the AB effect, a result which has made
possible the study of this phenomenon in perturbation theory.  Calculations
which have been carried out for spinless particles have found that unless an
additional (contact) interaction is introduced into the theory, divergences
similar to those encountered in quantum mechanical calculations will occur [6].

     An extension of the AB effect to include spin has also been carried out in
the context of the Dirac equation[7].  This has led to a recognition of the
fact that there must exist solutions of the wave equation which are singular
at the origin in the case that the spin orientation of the scattered particles
is such that the Zeeman interaction is attractive.  However, a remarkable
feature of the spin-1/2 case is the absence of divergences of the type which
characterize perturbation theory in the spinless case[8].  Clearly it would be
of interest to determine whether the corresponding spin-1/2 Galilean field
theory is also free of perturbative singularities.  It is in fact the goal of
the present work to demonstrate this result to fourth order in the coupling
(second order in $\alpha$).  

     In the following section a brief summary of the necessary field theoretic
tools is given, including a demonstration of how one proceeds from the
Hamiltonian of the field theory to a wave equation in the two particle sector. 
Section III introduces the fourth order diagram and carries out its evaluation
in the limit of very large mass for one of the two particles participating in
the scattering process.  Using this result it is possible to carry out in
section IV the evaluation of the scattering amplitude in the general mass case
and to determine the effect of the particle spins on the overall result.  Some
concluding remarks are offered in section V. 
\bigskip

\noindent II. Spin-1/2 Chern-Simons Galilean Field Theory

     Since the case of a spinless field in interaction with a Chern-Simons
field described by three components $\phi$ and $\phi_i$ (i=1,2) has been
discussed in some detail in ref. 5, it will be sufficient to present a somewhat
brief review of this subject, giving principal attention to those features
associated with the spin of the charged particle.  It is convenient to begin
with a single spin-1/2 particle of mass $M$.  As shown by L\'evy-Leblond [9] a
first order wave equation requires a four component spin-1/2 field operator in
three spatial dimensions.  For two spatial dimensions, however, a two component
field operator suffices for the description of a single spin component (just as
in the case of the Dirac equation in two spatial dimensions).  For the free
field case such an operator can be taken to satisfy the equation

\begin{equation}
\left[(1/2)(1+\sigma_3)i{\partial \over \partial t} +
i\mbox{\boldmath $\sigma$}
\cdot
\mbox{\boldmath $\nabla$} +
M(1-\sigma_3)\right]\psi=0
\end{equation}

\noindent where $\sigma_3$ is the usual third Pauli matrix while the matrices
$\sigma_i (i=1,2)$ are the set $(\sigma_1,s\sigma_2)$ where s is twice the
spin projection (+1 for spin \lq\lq up" and -1 for spin \lq\lq down").

     The combined system of interacting spinor and Chern-Simons field can be
described by the Lagrangian

\begin{eqnarray*}
{\cal L} & = & \psi^\dagger \left[ {1 \over 2}(1+ \sigma_3)
(i{\partial \over \partial t}-g \phi) 
+i \mbox{\boldmath $\sigma$} \cdot (\mbox{\boldmath $\nabla$}
-ig \mbox{\boldmath $\phi$}) + M(1- \sigma_3) \right]\psi 
 \\
& & \quad  -{1\over 2}\phi \mbox{\boldmath $\nabla$}
\times \mbox{\boldmath $\phi$} - 
{1 \over 2}\mbox{\boldmath $\phi$}
\times \mbox{\boldmath$\nabla$} \phi -
{1 \over 2}\mbox{\boldmath $\phi$}
 \times {\partial \over \partial t}
\mbox{\boldmath $\phi$}
\end{eqnarray*}

\noindent which implies the equations of motion

\begin{equation}
-\mbox{\boldmath $\nabla$} \times 
\mbox{\boldmath $\phi$} =g \rho 
\end{equation}

\begin{equation}
\epsilon_{ij}\left[{\partial \over \partial t} \phi_j + \nabla_j \phi
 \right] =g j_i 
\end{equation}

\begin{equation}
\left[ {1 \over 2}(1+ \sigma_3)(i{\partial \over \partial t} -g \phi)
+i\mbox{\boldmath $\sigma$}\cdot
(\mbox{\boldmath $\nabla$} -ig\mbox{\boldmath $\phi$})
+ M(1- \sigma_3) \right]\psi=0. 
\end{equation}

\noindent where 

$$\rho = \psi^\dagger{1 \over 2}(1+ \sigma_3) \psi$$

$$j_i = \psi^\dagger \sigma_i \psi.$$

\noindent In the radiation gauge 

       $$\mbox{\boldmath $\nabla$}\cdot\mbox{\boldmath $\phi$}=0,$$

\noindent Eq.(2) has the solution 

\begin{equation}
\phi_i = -g\epsilon_{ij}\nabla_j
\int d^2x^\prime\;  
 {\cal D}(x-x^\prime)\rho (x^\prime) 
\end{equation}                                          

\noindent where ${\cal D}(x)$ is defined by 

         $$-\mbox{\boldmath $\nabla^2$}{\cal D}(x) = \delta({\bf x})$$

\noindent or more explicitly by 

$${\cal D}(x)= -{1 \over 4\pi}\ell n {\bf x^2} + {\rm constant}.$$

\noindent Upon insertion of (5) into (3) one obtains for $\phi (x)$ the result

\begin{equation}
\phi (x)=g\int d^2x\;  {\bf j}(x')\times\mbox{\boldmath $\nabla$}
{\cal D}(x-x').
\end{equation}

\noindent It is worth emphasizing here the well known fact that there are no
independent degrees of freedom associated with the gauge fields $(\phi,
\phi_i)$ since they are expressible as explicit functions of the charged
spin-1/2 fields[4] as is clearly seen from Eqs.(5) and (6).  

     Also of interest is a more explicit display of the content of Eq.(4). 
Denoting the upper and lower components by $\varphi$ and $\chi$ respectively,
one has for $\varphi$ the result

\begin{equation}
(i{\partial\over\partial t} -g\phi)\varphi = (\Pi_1 - is\Pi_2)\chi 
\end{equation}

\noindent where $\Pi_i=-i\nabla_i -g\phi_i$.  Because of the presence of the
time derivative in (7) it is a true equation of motion as opposed to
the equation for $\chi$ which is of the form

         $$2M\chi = (\Pi_1 + is\Pi_2)\varphi.$$

\noindent Thus $\chi$ is a dependent field operator which is locally defined
in terms of $\varphi$ and the gauge field operators.

     Application of the action principle allows one to infer the equal time
anticommutation relation

\begin{equation}
\{\varphi (x),\varphi^\dagger(x^\prime)\}=\delta({\bf x-x^\prime})
\end{equation}

\noindent and the form of the conserved mass operator of the theory

$${\cal M}=M\int d^2x\;  \varphi^\dagger\varphi .$$

\noindent Since it will be convenient to consider the AB scattering of
dissimilar spin-1/2 particles in this work, the foregoing analysis will
henceforth be understood to include two fields $\psi$ and $\psi^\prime$ of
masses $M$ and $M^\prime$ respectively, each of which has identical coupling to
the Chern-Simons gauge field, and (possibly different) spin projections $s$ and
$s^\prime$.  Thus the commutation relation (8) is assumed to apply to the
fields $\varphi$ and $\varphi^\prime$ separately while the mass operator
becomes 

$${\cal M}=\int d^2x\; [M\varphi^\dagger\varphi +
M^\prime\varphi^{\prime\dagger}\varphi^\prime].$$

     The form of the interaction implies the existence of an additional
(global) symmetry which leads to the conclusion that each of the two terms in
${\cal M}$ is separately conserved.  This allows the states of the system to
be divided into sectors each of which is characterized by non-negative integers
$N$ and $N^\prime$ which denote the numbers of particles of masses $M$ and
$M^\prime$ respectively.  These states can be denoted by $|N,N^\prime>$ and
are constructed according to    

\begin{eqnarray*}
|N, N^\prime > & = & \int d^2x_1\ldots d^2x_Nd^2x_1^\prime 
\dots d^2x_{N^\prime}^\prime\; \varphi^\dagger (x_1)
\ldots\varphi^\dagger(x_N)\varphi^{\prime\dagger}(x_1^\prime)
 \\
\qquad & &
\ldots\varphi^{\prime\dagger}(x_{N^\prime}^\prime)f(x_1,
\ldots x_N;x_1^\prime,
\ldots x_{N^\prime}^\prime)|0>
\end{eqnarray*}

\noindent where $|0>$ is the vacuum or zero particle state

\begin{equation}
{\cal M}|0>=0
\end{equation}

\noindent and $f(x_1,\ldots x_N;x_1^\prime,\ldots x_{N^\prime}^\prime)$ is the
$N+N^\prime$ particle wave function.  The consistency of Eq.(9) clearly
requires that $\varphi$ and $\varphi^\prime$ annihilate the vacuum, i.e., 

                    $$\varphi (x)|0>=0$$

                 $$\varphi^\prime (x)|0>=0.$$

     The energy operator is inferred to have the form

$${\cal H}=\int d^2x\;  \left[{1\over
2M}\varphi^\dagger(\Pi_1+is\Pi_2)(\Pi_1-is\Pi_2)\varphi+{1\over2M^\prime}
\varphi^{\prime\dagger}(\Pi_1+is^\prime\Pi_2)(\Pi_1-is^\prime\Pi_2)
\varphi^\prime\right]$$

\noindent and allows the formulation of the eigenvalue equation

\begin{equation}
{\cal H}|N, N^\prime>=E|N,N^\prime>.
\end{equation}

\noindent One solves Eq.(10) by considering separately the various combinations
of $N$ and $N^\prime$.  Thus one clearly has the vacuum state for $N=N^\prime=0$
while the choices $N=1, N^\prime =0$ and $N=0,N^\prime=1$
yield the trivial results 
                    
                    $$(E+{1\over2M}\nabla^2)f({\bf x})=0$$

\noindent and 

              $$(E+{1\over2M^\prime}\nabla^{\prime 2})f({\bf x^\prime})=0$$

\noindent respectively.  The case $N=N^\prime=1$ is the sector which describes
the AB scattering of dissimilar fermions.  Application of Eq.(10) in this case
is found to imply the wave equation

\begin{eqnarray}
\lefteqn{ Ef({\bf x},{\bf x^\prime}) = } \nonumber \\
 &  & - \left\{ {1 \over 2M} \left[ \left[ 
\nabla_i + i{g^2\over 2\pi}\epsilon_{ij}
{(x-x^\prime)_j\over(x-x^\prime)^2} \right]^2 
- sg^2\delta({\bf x-x^\prime}) \right] \right. \nonumber \\
 &  & \left.
+{1\over 2M^\prime}\left[ \left[ \nabla_i-i{g^2\over 2\pi}\epsilon_{ij}
{(x-x^\prime)_j\over(x-x^\prime)^2} \right]^2
-s^\prime g^2\delta({\bf x-x^\prime})\right] \right\} f ({\bf x},{\bf x^\prime})
\end{eqnarray}

\noindent with similar results following for the cases $N=2, N^\prime=0$
 and $N=0, N^\prime=2$ 
which describe the AB scattering of identical particles of masses $M$
and $M^\prime$ respectively.  Worth noting in (11) is the explicit appearance
of the spin dependent terms proportional to $s$ and $s^\prime$.  These are of
the contact type and describe the Zeeman interaction of the magnetic moments of
the particles.   Since the system described by Eq.(11) allows one to
avoid the inessential complication of the Pauli principle and has the further
advantage of allowing the simultaneous consideration of parallel and
antiparallel spins, the remainder of this paper will focus exclusively on the
unequal mass case.  
                                                    
     One solves  Eq.(11) in the usual way by separation into the center-of-mass
coordinates and the relative coordinates ${\bf r}={\bf x-x^\prime}$. 
This leads to the reduced equation for the wave function $f({\bf r})$

\begin{equation}
\left[ \left[ \nabla_i-i\alpha
\epsilon_{ij}r_j/r^2 \right]^2-\alpha\mu({s\over M}+{s^\prime \over M^\prime})
{1\over r}\delta (r) +k^2 \right] f({\bf r})=0 
\end{equation}

\noindent where $k^2$ is the wave number in the center-of-mass frame, $\mu$ is
the reduced mass
                      $$\mu = {M M^\prime \over M+M^\prime},$$

\noindent and use has been made of the definition $g^2/2\pi=\alpha$.  (It is to
be noted that negative values of $\alpha$ can be obtained by consideration
of a Chern-Simons theory which is identical in all respects except for a change
in the sign of the terms in the Lagrangian which are quadratic in the gauge
fields.)  The result (12) in the case $s=s^\prime$ reduces to

$$
\left[ \left[ \nabla_i-i\alpha
\epsilon_{ij}r_j/r^2 \right]^2-\alpha s{1\over r}\delta(r)
+k^2 \right]f({\bf r})=0
$$

\noindent and is the basis for the quantum mechanical description of spin-1/2 AB
scattering [7].  Since it is in the unique case $s = s^\prime$ that
one has a divergence free perturbation expansion in quantum mechanics [8], the
goal of the remainder of this paper is to provide a corresponding demonstration
that to the one loop order it is only in this special case that the
corresponding field theoretic perturbation calculation is also divergence free.
\bigskip

\noindent III. Fourth Order AB Scattering

     In order to carry out the desired field theoretic perturbation expansion
it is necessary to prescribe the propagators and vertices associated with the
model.  The free field equation (1) for the field $\psi$ implies the momentum
space equation for the mass $M$ fermion propagator

$$\left[{1\over 2}(1+\sigma_3)E -\mbox{\boldmath$\sigma$}\cdot{\bf p}
+M(1-\sigma_3)\right]G({\bf p},E)=1$$

\noindent which has the solution

$$G({\bf p},E)=\left[M(1+\sigma_3) +\mbox{\boldmath$\sigma$}\cdot
{\bf p}+{1\over 2}(1-\sigma_3)E\right][2ME-p^2+i\epsilon]^{-1}.$$

\noindent  With respect to the gauge fields it is convenient to introduce a
notation such that $\phi^\alpha$ denotes the set ($\phi^0,\phi^i$) with
$\phi^0$ identified with $\phi$.  By standard means one then infers [4] that
the Chern-Simons propagator is given by 

$${\cal G}^{\alpha \beta}({\bf k})=i\epsilon^{\alpha\beta j}k_j{1\over k^2}.$$

\noindent Inspection of the form of ${\cal L}$ allows one to infer the fact that
the vertex matrices $\Gamma^\alpha$ are given by the set 
$({1\over 2}(1+\sigma_3),\sigma_i)$.

\begin{figure}[hp]
\epsffile[-40 520 250 600]{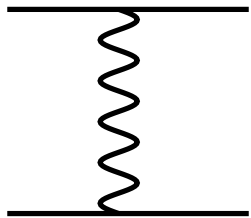} 
\caption{The second order scattering diagram}
\label{fig1}
\end{figure}

     To second order in $g$ one has only the single $\phi$ exchange diagram
displayed in Fig.1.  It is proportional to 

\begin{equation}
\left[{1\over 2}(1+\sigma_3)^{(1)}\sigma_i^{(2)}-{1\over 2}
(1+\sigma_3)^{(2)}\sigma_i^{(1)}\right]\epsilon_{ij}k_j{1\over k^2}
\end{equation}

\noindent where $k_i$ is the momentum transfer for the fermion of mass $M$ of
incoming momentum $p_i$ and outgoing momentum $p_i^\prime$.  A
superscript notation has been used to specify the matrices of the two
particles so that (1) and (2) refer respectively to mass $M$ and $M^\prime$
particles.  It is to be noted that (13) is to be evaluated between $u({\bf
p})^{(1)}u({\bf-p})^{(2)}$ and $u^*({\bf p^\prime})^{(1)}u^*(-{\bf
p^\prime})^{(2)}$ (working in the center-of-mass
frame) where the $u$'s are the relevant free particle spinors.  From (1) one
infers these to be of the form 

\begin{eqnarray*}
u({\bf p}) = \left( \begin{array}{c}
                   1\\
               {p_1+i sp_2\over 2M}
                     \end{array} \right).
\end{eqnarray*}  
                                    
\noindent Denoting the angle between {$\bf p$} and 
{$\bf p^\prime$} by $\theta$ one finds that
an evaluation of (13) between the indicated spinors yields a scattering
amplitude proportional to 

  $${g^2\over \mu\;\sin(\theta/2)}
  [\cos(\theta/2)-i\mu({s\over M}+{s^\prime\over
M^\prime})\sin(\theta/2)]$$

\noindent which reduces in the $s=s^\prime$ case to 

      $${g^2\over \mu\;\sin(\theta/2)}e^{-is\theta/2}$$

\noindent in agreement with the order $\alpha$ result which one obtains from an
expansion of the exact scattering amplitude [7].  Similar results have been
obtained to this order using covariant perturbation theory in the infinite
$M^\prime$ limit [10].

     Since the spin-1/2 scattering amplitude is known [7] to have no
$O(\alpha^2)$ corrections, one seeks to verify that the fourth order in $g$
result is both finite and null.  The specific diagram is displayed in Fig. 2
and is formally given by

\begin{eqnarray}
& \hbox{ }  & \int {dk\;dE\over (2\pi)^3}
\left\{ \Gamma^{\alpha}[M(1+\sigma_3)-\mbox{\boldmath$\sigma$}
\cdot
{\bf k}-{1\over2}E(1-\sigma_3)]\Gamma^\beta\right\}^{(1)}
\nonumber \\
& & \quad
\left\{ \Gamma^\kappa [M^\prime(1+\sigma_3)+\mbox{\boldmath $\sigma$}
\cdot 
{\bf k}+{1\over 2}(E+{p^2\over 2\mu})(1-\sigma_3)]\Gamma^\lambda\right\}^{(2)} 
\nonumber \\
& & \quad
{\cal G}_{\beta\lambda}({\bf p+k})
{\cal G}_{\alpha\kappa}({\bf p^\prime +k})
{1\over -2ME-k^2+i\epsilon} \;\;
{1\over 2M^\prime(E+{p^2\over 2\mu})-k^2+i\epsilon}.
\end{eqnarray}

\begin{figure}[htbp]
\epsffile[-20 520 250 600]{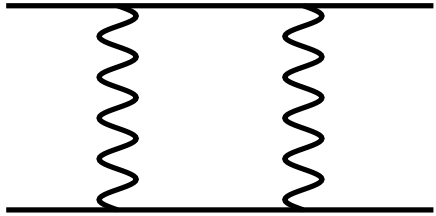} 
\caption{The fourth order scattering diagram}
\label{fig2}
\end{figure}

\noindent Although the integration over E is superficially linearly divergent,
in actuality the divergence is only a logarithmic one.  This can be seen by
noting that the antisymmetry of ${\cal G}_{\alpha\beta}$ in $\alpha$ and
$\beta$ implies that at least one of the factors of
 $E$ in the numerator of (14)
is necessarily multiplied by the product of 
$(1+\sigma_3)$ with $(1-\sigma_3)$ and
thus vanishes.  The remaining divergence must be regulated in a Galilean
invariant manner and requires some care.
    
     It will be convenient to classify contributions to (14) according
to whether the vertex indices are spatial ($\alpha=i$) or temporal
($\alpha=0$).  Referring to (14) one sees that the case $\alpha$ being a
temporal (spatial) index requires that $\kappa$ be a spatial (temporal) one
(and similarly for $\beta$ and $\lambda$).  Thus it follows that
(14) decomposes in a natural way according to whether a given fermion line has 
only temporal vertices, only spatial vertices, or mixed vertices.  Contribution
1 will thus be that part of (14) which has purely spatial vertices in
the mass $M$ propagator (temporal vertices in the mass $M^\prime$ propagator),
contribution 2 will be the corresponding case in which $M$ and $M^\prime$ are
exchanged, and contribution 3 will be the two parts in which mixed vertices
occur.  It is to be noted that there is no divergence in the $E$ integration for
contribution 3.

     In the limit of large $M^\prime$ only contribution 1 survives.  This is a
consequence of the fact that terms of the form 
$(1+\sigma_3)\sigma_iu({\bf p})^{(2)}$ 
vanish for large $M^\prime$.  It is worth noting that the surviving
term is precisely what one would consider in the case of scattering
from a fixed flux tube source.  Regularization is accomplished by  the
replacement 

$${1\over -2ME-k^2+i\epsilon}\rightarrow{1\over -2ME-k^2+i\epsilon}-
{1\over -2M(E+U)-k^2+i\epsilon}.$$

\noindent and subsequently taking the limit $U\rightarrow\infty$.  This is a
Galilean invariant regularization scheme since it consists of the
addition of an internal energy term to the Galilean invariant quantity in the
fermion propagator. Upon performing the integration over E one finds that (14)
reduces to 

\begin{eqnarray}
-i \int {dk\over (2\pi)^2}\;
{1\over({\bf p+k})^2}
{1\over ({\bf p^\prime +k})^2}
{1\over p^2-k^2+i\epsilon}
\mbox{\boldmath $\sigma$}
\times ({\bf p^\prime+k}) \nonumber \\
\left[ M(1+\sigma_3)-\mbox{\boldmath $\sigma$}\cdot
{\bf k}+{1\over 2}(1-\sigma_3)(p^2/2M)\right]
\mbox{\boldmath $\sigma$}\times({\bf p+k}),
\end{eqnarray}
which is seen by simple power counting to be finite.  Thus the AB scattering 
of spin-1/2 particles by a fixed flux tube is finite in this order.  This is to
be contrasted with the spin zero case which is, of course, rendered finite at
the one loop level only by the addition of a suitable contact term.

     It remains to be seen whether the result (15) vanishes on the \lq\lq
internal energy shell", i.e., when 

\begin{equation}
\left[{1\over 2}(1+\sigma_3)(p^2/2M) -\mbox{\boldmath$\sigma$}
\cdot {\bf p}+M(1-\sigma_3)\right]u({\bf p})=0.
\end{equation}
On applying (16) it is found that (15) becomes
\newpage

\begin{eqnarray*}
-i\sigma_i
\int {dk\over(2\pi)^2}
{1\over p^2-k^2+i\epsilon}
\left\{ {1\over 2}
\left[ {(p+k)_i\over({\bf p}+{\bf k})^2}+{(p^\prime+k)_i\over
({\bf p^\prime}+{\bf k})^2} \right] \right.\\
\; \qquad + \left.
 \epsilon_{ij}\left[(p^\prime+k)_j({\bf p}\times {\bf k})+
 (p+k)_j({\bf p^\prime} \times
{\bf k})\right]
{1\over({\bf p}+{\bf k})^2}{1\over ({\bf p^\prime} +{\bf k})^2}
\right\}.
\end{eqnarray*}

\noindent Symmetry considerations imply that the integral in this expression 
can be written in terms of two scalar functions $A({\bf p,p^\prime})$ and
$B({\bf p,p^\prime})$ as  

\begin{equation}
A({\bf p,p^\prime})(p+p^\prime)_i +
  B({\bf p,p^\prime})\epsilon_{ij}(p-p^\prime)_j({\bf p}\times{\bf p^\prime}).
\end{equation}

\noindent Since the matrix element of $\sigma_i$ is given by

$$u^*({\bf p^\prime})\sigma_iu({\bf p})={1\over 2M}\left[(p+p^\prime)_i
+is\epsilon_{ij}(p-p^\prime)_j\right],$$
\noindent it is necessary to contract (17) with $(p+p^\prime)_i$ and
$\epsilon_{ij}(p-p^\prime)_j$.  In the former case one obtains after
some algebra that 

\begin{eqnarray}
A({\bf p,p^\prime})({\bf p+p^\prime})^2 -2B({\bf p,p^\prime})({\bf p}
\times{\bf p^\prime})^2 &=& \int {dk\over (2\pi)^2}
\left\{
{4 ({\bf p}\times {\bf k}) ({\bf p^\prime}\times {\bf k})\over
({\bf p+k})^2({\bf p^\prime +k})^2}
{1\over p^2-k^2+i\epsilon} \right.\nonumber \\
&  & +  \left. 
{ ({\bf p}+{\bf k}) \cdot ({\bf p^\prime} +{\bf k})
\over ({\bf p}+{\bf k})^2
({\bf p^\prime +k)^2} } +
{1\over p^2-k^2+i\epsilon}\right \}.
\end{eqnarray}

\noindent Upon comparison with ref. 6 one sees that the result (18) plus the
corresponding expression with ${\bf p^\prime}\rightarrow{\bf -p^\prime}$ is
proportional to the scattering amplitude for identical spinless particles for
the case in which a contact term of the appropriate magnitude has been
included.  Since the latter was specifically constructed so as to give a
vanishing result, it is plausible that the right hand side of (18) also
vanishes.  This can in fact be verified by direct calculation.  Worth
emphasizing here is the fact that the calculation of ref. 6 required that the
noncovariant cutoffs of the $k$ integrals for both the Chern-Simons interaction
and the contact term be taken to be the same.  No such assumption is required
here since the integral in (15) is finite.

     To complete the argument it is also necessary to verify
that contraction of $\epsilon_{ij}(p-p^\prime)_j$ with (17) gives a vanishing
result.  Calculation shows that one again obtains the right hand side of (18) 
up to an overall kinematic factor, thereby establishing that the one loop
correction vanishes in the spin-1/2 case for $M^\prime\rightarrow\infty$.  The
removal of this latter condition is accomplished in the following section.  

\bigskip

\noindent IV The General Mass Case

\bigskip

     In order to consider the case of AB scattering with general masses $M$ and
$M^\prime$ one returns to (14).  Upon regulating the divergence in the energy
integration in contributions 1 and 2 as previously described one obtains
\begin{eqnarray}
{-i\over M + M^\prime} &  & \int
{dk\over (2\pi)^2}\; {1\over ({\bf p+k})^2}\; 
{1\over {\bf(p^\prime +k)^2}}\; {1\over p^2-k^2+i\epsilon} \nonumber \\
& & \!\!\!\!\!\!\!\!\!\!\!\!\!\!\!\!\!\!
\Biggl\{ 
M^\prime
\Biggl[ \mbox{\boldmath$\sigma$}\times ({\bf p^\prime+k})
\Bigl[
M(1+\sigma_3)-\mbox{\boldmath $\sigma$} \cdot
{\bf k}+{1\over2}(1-\sigma_3) \left( {p^2\over 2\mu}-
{k^2\over 2M^\prime} \right) 
\Bigr] \mbox{\boldmath $\sigma$}\times ({\bf p+k}) \Biggr]^{(1)}
  \nonumber \\
& &  \!\!\!\!\!\!\!\!\!\!\!\!\!\!\!\!\!\!
{1\over 2}(1+\sigma_3)^{(2)} +M{1\over 2}(1+\sigma_3)^{(1)}
\Biggl[ \mbox{\boldmath $\sigma$}\times({\bf p^\prime +k})
\Bigl[ M^\prime (1+\sigma_3)+ \mbox{\boldmath$\sigma$}\cdot 
{\bf k}+\left({p^2\over 2\mu}-
{k^2\over 2M}\right) \Bigr] 
\nonumber \\
& & \!\!\!\!\!\!\!\!\!\!\!\!\!\!\!\!\!\!
\mbox{\boldmath$\sigma$}\times({\bf p+k})\Biggr]^{(2)}
-{1\over 2} 
\Biggl[ {1\over2}(1+\sigma_3)
\Bigl[ M(1+\sigma_3)-\mbox{\boldmath$\sigma$}\cdot {\bf k}
\Bigr] \mbox{\boldmath$\sigma$}\times ({\bf p+k})
\Biggr]^{(1)}
 \nonumber \\
& & \!\!\!\!\!\!\!\!\!\!\!\!\!\!\!\!\!\! 
\Biggl[ \mbox{\boldmath $\sigma$}\times ({\bf p^\prime +k})
\Bigl[ M^\prime(1+\sigma_3)+\mbox{\boldmath $\sigma$}\cdot {\bf k} \Bigr]
{1\over 2}(1+\sigma_3) 
\Biggr]^{(2)} \nonumber \\
& & \!\!\!\!\!\!\!\!\!\!\!\!\!\!\!\! 
-{1\over 2}\Biggl[ \mbox{\boldmath$\sigma$}\times ({\bf p^\prime +k}) 
\Bigl[ M(1+\sigma_3) -\mbox{\boldmath $\sigma$} \cdot {\bf k}
\Bigr] {1\over 2}(1+\sigma_3) 
\Biggr]^{(1)} \nonumber \\
& & \!\!\!\!\!\!\!\!\!\!\!\!\!\!\!\!\!\!
\Biggl[ {1\over 2}(1+\sigma_3)
\Bigl[ M^\prime (1+\sigma_3)+\mbox{\boldmath $\sigma$}\cdot {\bf k}
\Bigr] \mbox{\boldmath $\sigma$}\times ({\bf p+k}) 
\Biggr]^{(2)}
\Biggr\}. 
\end{eqnarray}
 
\noindent The integrand of this expression contains four separate terms, of
which the first is the type 1 contribution, the second is the type 2, and the
last two are the (mixed) type 3 contribution.  Considerable simplification of
the former is achieved by applying the condition (16) and the corresponding one
for the mass $M^\prime$ part together with the result established in the
preceding section concerning the vanishing of the type 1 (type 2) contribution
in the infinite $M^\prime$ (infinite $M$) limit.  One finds that the type 1 and
type 2 contributions to the curly bracket in Eq.(19) thereby reduce to

\begin{eqnarray*}
(p^2-k^2){1\over 2}(1+\sigma_3)^{(1)}{1\over 2}(1+\sigma_3)^{(2)}
\;\;\;\;\;\;\;\;\;\;\;\;\;\;\;\;\;\;\;\;\;\;\;\;
\\
{1\over 2}\left\{\left[\mbox{\boldmath $\sigma$}
\times({\bf p^\prime +k})\mbox{\boldmath
$\sigma$}\times({\bf p+k})\right]^{(1)} + \left[\mbox{\boldmath $\sigma$}\times
({\bf p^\prime +k})\mbox{\boldmath $\sigma$}\times 
({\bf p+k})\right]^{(2)}
\right\}
\end{eqnarray*}

\noindent which can be written as

$$(p^2-k^2){1\over 2}(1+\sigma_3)^{(1)}{1\over 2}(1+\sigma_3)^{(2)}
\left[({\bf p+k})\cdot({\bf p^\prime +k})
 - {i\over 2}(s+s^\prime)({\bf p+k})\times
({\bf p^\prime +k})\right].$$
                         
     The type 3 contributions also undergo considerable simplification when
(16) is invoked.  One finds that these reduce to

\begin{eqnarray*}
-{1\over 2}\;{1\over 2}(1+\sigma_3)^{(1)}{1\over 2}(1+\sigma_3)^{(2)}
\Bigl\{
\bigl[ 2{\bf k}\times{\bf p}-is({\bf p+k})^2\bigr]^2
\bigl[
2{\bf p^\prime}\times{\bf k}-is^\prime({\bf p^\prime +k})^2\bigr]
 \\
\qquad\qquad  +
\bigl[2{\bf k}\times{\bf p}-is^\prime({\bf p+k})^2\bigr]
\bigl[2{\bf p^\prime}\times {\bf k}-is({\bf p^\prime+k})^2\bigr]\Bigr\},
\end{eqnarray*}
which can be written as

\begin{eqnarray*}
&\mbox{} & {1\over 2}(1+\sigma_3)^{(1)}{1\over 2}(1+\sigma_3)^{(2)}
 \Bigl\{ 4 ({\bf p}  \times  {\bf k})
 ({\bf p^\prime}\times{\bf k}) + ss^\prime ({\bf p+k})^2 
({\bf p^\prime+k})^2 \nonumber \\
&  &  - i(s+s^\prime) 
\bigl[ ({\bf k}\times{\bf p^\prime})({\bf p+k})^2
- ({\bf k}\times {\bf p})({\bf p^\prime}+{\bf k})^2
\bigr]
 \Bigr\}.
\end{eqnarray*}
            
\noindent It is seen by inspection that the terms linear in $s$ and $s^\prime$
vanish upon doing the angular integration so that one obtains upon combining
all these results the reduction of (19) to

\begin{eqnarray*}
-{i\over M+M^\prime} {1\over 2}(1+\sigma_3)^{(1)}{1\over 2}
(1+\sigma_3)^{(2)}
\int {dk\over (2\pi)^2} 
 \left\{  
{4({\bf p}\times {\bf k})({\bf
p^\prime}\times{\bf k})\over ({\bf p+k})^2({\bf p^\prime + k})^2}
{1\over p^2-k^2+i\epsilon} \right.\\
\left. + {ss^\prime\over p^2-k^2+i\epsilon}+ 
{ ({\bf p+k}) \cdot ({\bf p^\prime +k})
\over ({\bf p+k})^2 ({\bf p^\prime +k})^2 }
-{i\over 2}(s+s^\prime) { ({\bf p+k})\times({\bf p^\prime +k})
\over ({\bf p+k})^2 ({\bf p^\prime +k})^2}
\right\}.
\end{eqnarray*}

\noindent Again it can be shown that the term which is linear in $s+s^\prime$
vanishes upon performing the angular integration.  This leaves one with a
result which is remarkably similar to (18).  Since the latter is known to
vanish, one readily finds that the one loop correction to spin-1/2 AB
scattering is given by 

$${-i\over M+M^\prime}
{1\over 2}(1+\sigma_3)^{(1)}
{1\over 2}(1+\sigma_3)^{(2)}
\int {dk\over (2\pi)^2}
{(ss^\prime-1)\over (p^2-k^2+i\epsilon)}$$

\noindent which is the final result of this calculation.  For $s s^\prime=1$
(i.e., parallel spins) the result is thus seen to be both finite and vanishing
to order $\alpha^2$.  In this case one has equivalence to the $s=s^\prime$
limit of Eq.(12) which is known [8] to have a divergence free perturbative
expansion.  Conversely, for the antiparallel spin configuration one has either a
magnetic moment which vanishes ($M=M^\prime$) or one which has a g-factor less
than 2 (general $M,M^\prime$) and is thus not equivalent to the spin-1/2 AB
scattering system studied in ref. 7.  In each of these latter cases one expects
divergences in perturbative calculations in agreement with the results obtained
here.

\bigskip

\noindent V Conclusion

     This work has succeeded in restoring a certain symmetry between spin zero
and spin-1/2 work on the AB effect.  While the quantum mechanical AB effect had
been solved for both the scalar [1] and spinor [7] cases and their perturbative
expansions studied in both applications [2,3,8], only the scalar theory had been
studied previously as a perturbation expansion in field theory.  Although the
technical complications associated with the matrix algebra are quite
significant in the spin-1/2 field theory, it has in fact been found possible to
carry through the calculation of the AB scattering amplitude to fourth order in
the coupling constant and thereby reestablish the aforementioned balance
between the scalar and spinor theories. 

     It is certainly gratifying that the results of this study conform with
those which have been found in ref.8.  Beyond that, however, is the very useful
set of rewards which have followed from the use of nonidentical particles in
this study.  First of all it allowed one to begin the calculation with the much
more manageable examination of AB scattering from a fixed flux tube (i.e., the
$M^\prime\rightarrow\infty$ limit), obtaining in the process a result which
greatly facilitated the treatment of the more general case. Second, it allowed
one to avoid the extraneous complication arising from the Pauli principle.
Finally, and perhaps most significant of all, it allowed the simultaneous
consideration of the parallel and antiparallel spin cases.  It has been found
that in the former case the result is both finite and vanishing at the fourth
order while in the latter one encounters the divergences known to characterize
the spin zero theory.  All of these results are in conformity with calculations
which have been carried out in the context of quantum mechanics. 

     Finally, mention should be made of the fact that it must be regarded as
encouraging that calculations such as those presented here can be effectively
carried out.  This is a significant point since the spin-1/2 theory has a
crucial advantage over the corresponding scalar theory by virtue of its being
finite in perturbation theory without the {\em ad hoc} inclusion of additional
coupling terms.  As has already been mentioned, the fact that the spin-1/2
theory already implies such contact terms through the magnetic moment
interaction serves to eliminate ambiguities in the regularization of divergent
integrals.  It may thus be possible that such features will cause spin-1/2
perturbative calculations to see greater application in the future. 
     
\bigskip

This work is supported in part by the U.S. Department of Energy Grant No.
DE-FG02-91ER40685.

\medskip

\noindent References

\begin{enumerate}
\item Y. Aharonov and D. Bohm, Phys. Rev.{\bf 115}, 485 (1959).

\item E. L. Feinberg, Usp. Fiz. Nauk. {\bf 78}, 53 (1962) [Sov. Phys. Usp. 
{\bf5},753 (1963)]; E. Corinaldesi and F. Rafeli, Am. J. Phys. {\bf 46} 1185 
(1978).

\item Y. Aharonov, C. K. Au, E. C. Lerner, and J. Q. Liang, 
Phys. Rev. D{\bf 29},2396 (1984).

\item C. R. Hagen, Ann. Phys. (N.Y.) {\bf 157}, 342 (1984).

\item C. R. Hagen, Phys. Rev. D{\bf 31},848 (1985).

\item O. Bergman and G. Lozano, Ann. Phys. (N.Y.) {\bf 229}, 416 (1994);
 D. Bak and O. Bergman, Phys. Rev. D{\bf 51}, 1994 (1995).

\item C. R. Hagen, Phys. Rev. Lett. {\bf 64}, 503 (1990).

\item C. R. Hagen, Phys. Rev. D{\bf 52}, 2466 (1995).

\item J. M. L\'evy-Leblond, Commun. Math. Phys. {\bf 6}, 286 (1967).

\item F. Vera and I. Schmidt, Phys. Rev. D{\bf 42}, 3591 (1990). 

\end{enumerate}

\end{document}